\documentclass[final,epj]{svjour}
\usepackage{graphics}
\usepackage{amsmath,amssymb}

\begin{document}

\title{Investigation on Vibrational, Optical and Structural Properties of an Amorphous Se$_{0.80}$S$_{0.20}$ Alloy 
Produced by Mechanical Alloying}

\author{K. D. Machado \inst{1} \and D. F. Sanchez \inst{1} \and A. S. Dubiel \inst{1} 
\and S. F. Brunatto \inst{2} \and S. F. Stolf \inst{3} \and P. J\'ov\'ari \inst{4}}

\titlerunning{Vibrational, Optical and Structural Properties of {\em a}-Se$_{0.80}$S$_{0.20}$ Produced by MA}

\authorrunning{K. D. Machado et al.}

\institute{Departamento de F\'{\i}sica, Centro Polit\'ecnico, Universidade Federal do Paran\'a, 
81531-990, Curitiba, Paran\'a, Brazil 
\and Departamento de Engenharia Mec\^anica, Centro Polit\'ecnico, Universidade Federal do Paran\'a, 
81531-990, Curitiba, Paran\'a, Brazil 
\and Centro de Engenharia e Ci\^encias Exatas, UNIOESTE, 85903-000, Toledo, Paran\'a, Brazil
\and Hungarian Acad Sci, Res. Inst. Solid State Phys \& Opt, POB 49, H-1525 Budapest, Hungary
}

\offprints{K. D. Machado}
\mail{kleber@fisica.ufpr.br}

\abstract{
An amorphous Se$_{0.80}$S$_{0.20}$ alloy produced by Mechanical Alloying was 
studied by Raman spectroscopy, x-ray diffraction, extended x-ray absorption fine structure (EXAFS) and optical absorption 
spectroscopy, and also through reverse Monte Carlo simulations of its total structure factor and EXAFS data. Its 
vibrational modes, optical gap and structural properties as average interatomic distances and average coordination 
numbers were determined and compared to those found for an amorphous Se$_{0.90}$S$_{0.10}$ alloy. The results indicate 
that coordination numbers, interatomic distances and also the gap energy depend on the sulphur concentration.
}

\maketitle

\section{Introduction}

Owning to their technological applications in optoelectronic, optical, electronic  and memory 
switching devices the research on chalcogenide glasses formed by elements Se, S and Te has increased in recent years. 
Usually, chalcogenide alloys based on selenium have high transparency in the broad middle and far 
infrared region and have strong non-linear properties. Alloys formed by Se and S exhibits an electronic 
conductivity of p-type semiconductor and some properties related to these alloys were already studied. 
Thermal 
and electrical properties for amorphous samples were determined in some studies \cite{Kotkata1,Kotkata2,Mahmoud}, 
and recently Musahwar {\em et al}. \cite{Musahwar} measured dielectric and electrical properties for Se$_{x}$S$_{1-x}$ 
glasses ($x=1,0.95,0.90,0.85,0.80$), and the optical gap of Se$_x$S$_{1-x}$ thin films ($x=0.9,0.8,0.7,0.6$) 
were determined by Rafea and Farag \cite{Rafea}.
Ward 
\cite{Ward} obtained vibrational modes for some crystalline Se-S alloys through Raman spectroscopy.  
Concerning structural properties, relatively few investigations have 
been done on these alloys. The compositional range for fabrication of amorphous Se-S alloys was studied 
by Kotkata {\em et al}. \cite{Kotkata3,Kotkata4} through x-ray diffraction (XRD) measurements. They produced 
several amorphous and crystalline Se$_x$S$_{1-x}$ alloys and obtained some general results, 
indicating that amorphous alloys can be prepared in a relatively wide compositional range, from $x=0.5$ to 
$x=1$, the density of the alloys decreases as the S content is increased and the average total coordination 
number for Se atoms ($\langle N^{\text{Se}} \rangle =\langle N^{\text{Se-Se}} \rangle+ \langle N^{\text{Se-S}}\rangle$) is 2. 
Heiba {\em et al}. \cite{Heiba} also used 
the XRD technique to study the crystallization of Se$_x$S alloys, $x=20,30,40$, and crystallite sizes 
were determined from a refinement using the program FULLPROOF. A more detailed investigation of amorphous alloys 
was made by Shama \cite{Shama}, which reports the results obtained from simulations of the first shell of 
the radial distribution functions (RDF) determined for three amorphous Se$_x$S alloys, $x=10,30,40$. 
For these alloys, average coordination numbers and average interatomic distances for Se-Se pairs were determined, but 
the existence of Se-S pairs was not considered. 
The values found were $\langle N^{\text{Se-Se}}\rangle =2.10$ and $\langle r^{\text{Se-Se}}\rangle=2.369$~\AA, 
for Se$_{10}$S, and $\langle N^{\text{Se-Se}} \rangle =2.14$ and $\langle r^{\text{Se-Se}}\rangle =2.368$~\AA\ 
for Se$_{30}$S. It is interesting to note that all samples cited  were prepared by quenching, 
except those in \cite{Rafea}, which were prepared by thermal evaporation. Fukunaga 
{\em et al}. \cite{Fukunaga2} used the Mechanical Alloying (MA) \cite{MASuryanarayana} technique to produce amorphous 
Se$_x$S$_{1-x}$ alloys, in the compositions $x=1,0.90,0.80,0.70,0.60$, which were investigated using neutron diffraction (ND) 
followed by a RDF analysis. However, results were given only for 
the composition Se$_{0.60}$S$_{0.40}$, and, in this case, they found $\langle N^{\text{Se-Se}}\rangle =1.78\pm 0.018$ and 
$\langle r^{\text{Se-Se}}\rangle =2.37 \pm 0.002$ 
\AA, and again the possibility of having Se-S pairs was not taken into account. In addition, they assumed that 
the MA process only mixed Se chains and S rings, and there is no alloying at the atomic level.


Due to the promising applications of Se-S alloys and the lack of a sistematic investigation of their structures, 
we started a more detailed structural study about this alloys. In a recent article \cite{klese90s10} we investigated 
an amorphous Se$_{0.90}$S$_{0.10}$ alloy ({\em a}-Se$_{0.90}$S$_{0.10}$) produced by MA. We obtained vibrational 
properties, by Raman spectroscopy, optical properties, by optical absorption spectroscopy, and structural properties 
considering extended x-ray absorption fine structure spectroscopy (EXAFS) and synchrotron XRD. We also made reverse 
Monte Carlo (RMC) simulations \cite{rmcreview,rmc++} of the total structure factor ${\mathcal S}(K)$ obtained from the XRD 
data.

In the present study, we analized the formation of an amorphous Se$_{0.80}$S$_{0.20}$ alloy 
({\em a}-Se$_{0.80}$S$_{0.20}$) also produced by MA, and 
investigated its vibrational modes using Raman spectroscopy, its structural properties considering 
RMC simulations using both XRD ${\mathcal S}(K)$ and EXAFS $\chi(k)$ signal at Se K edge as input data and we also 
determined the optical gap by optical absorption spectroscopy. All data were compared to those obtained for 
{\em a}-Se$_{0.90}$S$_{0.10}$ and some interesting features can be inferred from the data obtained.

\section{Experimental Procedures}

Amorphous Se$_{0.80}$S$_{0.20}$ samples were produced by 
milling Se (Aldrich, purity $>$ 99.99\%) and S (Vetec, purity $>$ 99.5\%)  powders 
in the composition above. The powders were sealed together with 15 steel balls (diameter 10 mm), under 
argon atmosphere, in a steel vial. The weight ratio of the ball to powder was 9:1. The vial was 
mounted in a Fritsch Pulverisette 5 planetary ball mill and milled at 350 rpm. In order to keep 
the vial temperature close to room temperature, the milling was performed considering cycles of 20 min of 
effective milling followed by 10 min of rest. To investigate the formation of the alloy, XRD measurements 
were taken after 11 h, 23 h, 34 h, 46 h and 57 h of milling. They were done in a 
Shimadzu difractometer using Cu K$_\alpha$ radiation ($\lambda = 1.5418$ \AA) in a $\theta$--$2\theta$ scanning mode, 
considering a $\Delta 2 \theta$ step of 0.04$^\circ$ and each $2\theta$ point was measured during 22 s for all milling 
times except for 11 h, which was measured considering 1 s per point. 
After 57 h of milling, the XRD pattern was characteristic of amorphous 
samples, showing large amorphous halos without crystalline peaks.

Micro-Raman measurements were performed with a Renishaw spectrometer coupled to an optical microscope and a cooled 
CCD detector. The 6238 \AA\ line of an HeNe laser was used as exciting light, always in backscattering geometry. 
The output power of the laser was kept at about 1--3 mW to avoid overheating samples. All Raman measurements were 
performed with the samples at room temperature.

EXAFS measurements at Se K edge were taken at room temperatures in the transmission mode at beam line D08B - XAFS2 of 
the Brazilian Synchrotron Light Laboratory  - LNLS (Campinas, Brazil). Three ionization chambers were used as detectors. 
The {\em a}-Se$_{0.80}$S$_{0.20}$ sample was formed by placing the powder onto a porous membrane (Millipore, 0.2 $\mu$m 
pore size) in order to achieve optimal thickness (about 50 $\mu$m) and it was placed between the first and second 
chambers. A crystalline Se foil used as energy reference was placed between the second and third chambers. The beam 
size at the sample was 3 $\times$ 1 mm. The energy and average current of the storage ring were 1.37 GeV and 190 mA, 
respectively.



In order to obtain the EXAFS $\chi(k)$ signal on Se K edge to be used on RMC simulations, the raw EXAFS data were 
analyzed following standard procedures. EXAFS spectra were energy calibrated, aligned, and isolated from raw absorbance 
performing a background removal using the AUTOBK algorithm of the ATHENA \cite{athena} program, and the $\chi(k)$ 
signal was obtained.

Synchrotron XRD measurements were carried out at the BW5 beamline \cite{Palref} at HASYLAB. All data were 
taken at room temperature using a Si (111) monochromator and a Ge solid state detector. The energy of the incident 
beam was 121.3 keV ($\lambda  = 0.102$ \AA), and it was calibrated using a LaB$_6$ standard sample. The uncertainty 
in the wavelength is less than 0.5\%. The cross section of the beam was $1 \times 4$ mm$^2$ (h x v). 
Powder sample was filled into a thin walled (10 $\mu$m) quartz capillary with 2 mm diameter.
The energy and average current of the storage ring were 4.4 GeV and 
110 mA, respectively. Raw intensity was corrected for deadtime, background, 
polarization, detector solid angle and Compton-scattering as described in \cite{Palref}. 
The total structure factor was computed from the normalized intensity $I_a(K)$ according to Faber and Ziman 
\cite{Faber} (see eq. \ref{eqstructurefactor1}).

Absorbance measurements were carried out in a Shimadzu UV-2401-PC 
spectrometer. In these measurements the {\em a}-Se$_{0.80}$S$_{0.20}$ sample was mixed to KBr and pressed in the 
form of a pellet. KBr was used as support and reference.

\section{Theoretical Background}

\subsection{Structure Factors and RMC Simulations}

According to Faber and Ziman \cite{Faber}, the total structure factor ${\mathcal S}(K)$ is obtained 
from the scattered intensity per atom $I_a(K)$ through

\begin{eqnarray}
{\mathcal S}(K) &=& \frac{I_a(K)-\bigl[\langle f^2(K)\rangle - \langle f(K)\rangle^2
\bigr]}{\langle f(K)\rangle^2} \label{eqstructurefactor1}\,,\\
{\mathcal S}(K)&=& \sum_{i=1}^n{\sum_{j=1}^n{w_{ij}(K) {\mathcal S}_{ij}(K) }}\,,
\label{eqstructurefactor}
\end{eqnarray}

\noindent where $K$ is the transferred momentum,  
${\mathcal S}_{ij}(K)$ are the partial structure factors and $w_{ij}(K)$ are given by

\begin{equation}
w_{ij}(K) = \frac{c_i c_j f_i(K) f_j(K)}{\langle f(K)\rangle^2}\,,
\label{eqw}
\end{equation}
 
\noindent and

\begin{eqnarray*}
\langle f^2(K) \rangle &=& \sum_{i}{ c_i f_i^2(K)}\,,\\
\langle f(K) \rangle^2 &=& \Bigl[\sum_{i}{ c_i f_i(K)}\Bigr]^2 \,.
\end{eqnarray*}

\noindent Here, $f_i(K)$ is the atomic scattering factor 
and $c_i$ is the concentration of atoms of type $i$. The partial  
distribution functions $g_{ij}(r)$ are related to ${\mathcal S}_{ij}(K)$ and 
${\mathcal S}(K)$ through

\begin{equation}
g_{ij}(r) = \frac{2}{\pi} \int_0^{\infty}{K\bigl[{\mathcal S}_{ij}(K)-1 \bigr] 
\sin (Kr)\, dK}\,,
\end{equation}

\noindent and, from these functions, interatomic distances and coordination numbers 
can be determined. 

The structure factors defined by eq. \ref{eqstructurefactor1} can be used in RMC simulations.\ 
The algorithm of the standard RMC method are described 
elsewhere \cite{rmcreview,rmc++} and its application to 
different materials is reported in the literature 
\cite{nitikleber,Pal1,rmcNDXRD,Iparraguirre}. 

When using XRD data only, the idea is to minimize the function

\begin{equation}
\psi^2_{\text{XRD}} = 
\sum_{i=1}^m{\frac{\bigl[{\mathcal S}^{\rm{RMC}}(K_i) - {\mathcal S}(K_i)\bigr]^2}{\delta^2_{\text{XRD}}} }
\label{eqchi}
\end{equation}

\noindent where ${\mathcal S}(K)$ is the experimental total structure factor, ${\mathcal S}^{\rm{RMC}}(K)$ 
is the estimate of ${\mathcal S}(K)$ obtained by RMC simulations, $\delta_{\text{XRD}}$ is a parameter related 
to the convergence of the simulations and to the experimental errors and the 
sum is over $m$ experimental points. In our case, we added EXAFS data to the simulations, and the function 

\begin{equation}
\psi^2 = \psi^2_{\text{XRD}} + \psi^2_{\text{EXAFS}}
\label{eqchitotal}
\end{equation}

\noindent where

\begin{equation}
\psi^2_{\text{EXAFS}} = 
\sum_{i=1}^m{\frac{\bigl[\chi^{\rm{RMC}}(k_i) - \chi(k_i)\bigr]^2}{\delta^2_{\text{EXAFS}}} }
\label{eqchiexafs}
\end{equation}

\noindent is the function to be minimized. In eq. \ref{eqchiexafs}, $\chi(k_i)$ is the experimental EXAFS signal, 
$\chi^{\rm{RMC}}(k)$ is its estimate obtained using RMC simulations and $\delta_{\text{EXAFS}}$ is the parameter 
related to the experimental errors in the EXAFS signal. To perform the simulations we have considered the RMC program 
available on the Internet \cite{rmc++} and cubic cells with 16000 atoms. The total structure factor obtained 
from XRD measurements and the  EXAFS data at Se K edge were used as input data for the simulations. 

\subsection{Optical Band Gap Determination}
\label{secoptical}

To obtain the optical gap a simple and direct procedure can be used, and consists in determining 
the wavelength at which the extrapolations of the baseline and the absorption edge cross \cite{Boldish}. 
A more sofisticated approach makes use of a McLean analysis \cite{McLean} of the absorption edge, which furnishes 
more information about the lowest energy interband transition. The absorption coefficient $\alpha$ follows the 
equation 

\begin{equation}
\alpha h \nu = (h \nu - E_g)^{\frac{1}{n}}
\label{absorcao}
\end{equation}

\noindent where $E_g$ is the gap energy and $\nu$ is the frequency of the incident beam. The analysis consists of 
fitting the absorption edge to eq. \ref{absorcao} and determining experimental values for $E_g$ and $n$. $n=2$ 
correponds to a direct allowed transition. $n = \frac{2}{3}$ implies a direct  forbidden transition. $n = \frac{1}{2}$ 
is associated 
with an indirect allowed transition and $n = \frac{1}{3}$ implies an indirect forbidden transition. The absorbance 
$A$, the absorption coefficient $\alpha$ and the thickness $d$ of a sample are related by $\alpha = A/d$ 
and, for absorbance measurements on powders, in which the polycrystalline or amorphous samples are dispersed into a 
powder support such as KBr, the mixture is pressed in the form of pellet. In this case, the thickness $d$ and 
the absorption coefficient $\alpha$ of the sample become unknown. Thus, eq. \ref{absorcao} must be modified to 

\begin{equation}
A h \nu = C (h \nu - E_g)^{\frac{1}{n}}
\label{absorcao2}
\end{equation}

\noindent where $C$ represents the thickness of the sample and is a parameter to be included in the fitting 
procedure.

\section{Results and Discussion}
\label{secr}

\subsection{Formation of {\em a}-Se$_{0.80}$S$_{0.20}$}

\begin{figure}[h]
\begin{center}
\includegraphics{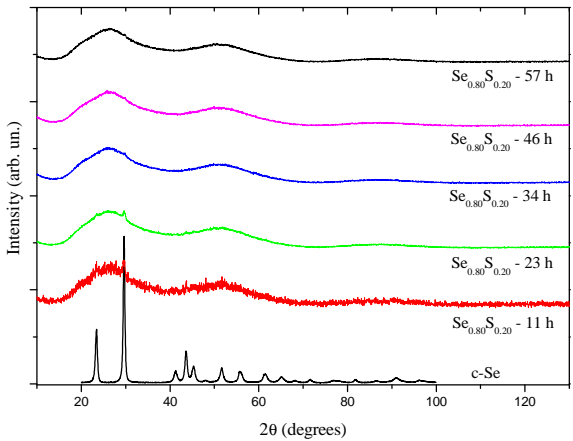}
\end{center}
\caption{\label{fig1} XRD measurements obtained for {\em a}-Se$_{0.90}$S$_{0.10}$ after selected milling times: 
11 h, 23 h, 34 h, 46 h and 57 h. The XRD pattern for {\em c}-Se is also shown for comparison.}
\end{figure}

In order to compare the evolution with milling time of {\em a}-Se$_{0.80}$S$_{0.20}$ and 
{\em a}-Se$_{0.90}$S$_{0.10}$ we stopped the milling at some chosen times and made conventional XRD measurements 
to verify the changes on the structure of the alloy. Fig. \ref{fig1} shows these XRD measurements obtained for 
{\em a}-Se$_{0.80}$S$_{0.20}$. In this figure the XRD pattern for the crystalline selenium powder ({\em c}-Se) used 
as the starting material is also shown for comparison. 

We decided to make the first stop after 11 h of milling. Considering the results obtained for 
{\em a}-Se$_{0.90}$S$_{0.10}$ (fig. 1 of Ref. \cite{klese90s10}), we supposed that at this time an intermediate 
structural state between those found at 4 h and at 14 h of milling would be found and, in fact, the XRD pattern 
for {\em a}-Se$_{0.80}$S$_{0.20}$ at 11 h of milling corroborates this assumption. 

The next stop was at 23 h, and the XRD pattern shows only the most intense crystalline Se peak around 30$^\circ$, 
besides the amorphous halos, being similar to the XRD pattern at 25 h of milling for {\em a}-Se$_{0.90}$S$_{0.10}$. 
Then we decided to make more stops, to verify the evolution of the remaining Se peak. At 34 h this peak was still 
seen in the XRD pattern, and disappeared only after 46 h of milling. After that, the amorphous structure is stable, 
as the XRD pattern at 57 h indicates.

It is interesting to note that, according Fukunaga {\em et al}. \cite{Fukunaga2}, after 
30 h of milling the {\em a}-Se$_{0.80}$S$_{0.20}$ alloy should be ready. 
Since MA is very sensitive to the milling conditions \cite{MASuryanarayana}, and considering that 
we used a different number of balls and a different total mass of the starting powders, 
resulting in a free volume inside the vial larger to us than to them, we believe 
these points explain the difference in the milling time needed to produce the alloy.

\subsection{Raman Spectroscopy Results}

\begin{figure}[h]
\begin{center}
\includegraphics{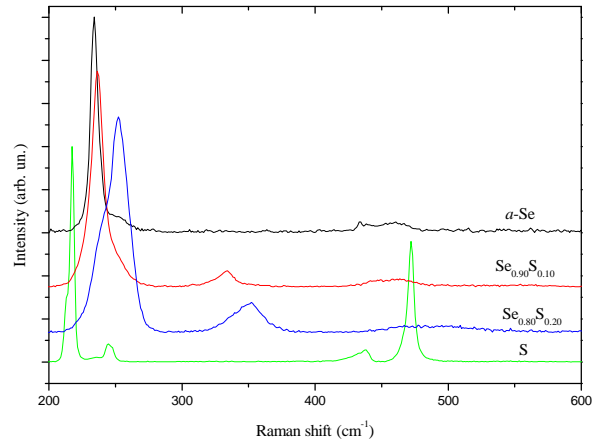}
\end{center}
\caption{\label{fig2} Raman spectra for {\em a}-Se$_{0.80}$S$_{0.20}$, {\em a}-Se$_{0.90}$S$_{0.10}$, {\em a}-Se 
and {\em c}-S.}
\end{figure}

In their study about Se-S amorphous alloys, Fukunaga {\em et al}. \cite{Fukunaga2} stated that MA produces only a 
mixing of Se chains and S rings and not a true alloy at the atomic level. This idea was also assumed by 
Shama \cite{Shama}, for a quenched sample. However, we have demonstrated that this assumption is wrong for 
{\em a}-Se$_{0.90}$S$_{0.10}$ \cite{klese90s10}, since some Raman modes related to Se-S vibrations appear in the 
Raman spectrum of this alloy. Thus, we expected that for {\em a}-Se$_{0.80}$S$_{0.20}$ these vibrations would also 
appear, besides some shifts on the other modes associated with Se-Se vibrations. Fig. \ref{fig2} shows the results 
obtained for this alloy, for amorphous Se, for {\em c}-S and also for {\em a}-Se$_{0.90}$S$_{0.10}$, for comparison. 

Fig. \ref{fig2} shows clearly that some vibrational modes associated with {\em a}-Se appear 
in {\em a}-Se$_{0.90}$S$_{0.10}$ and {\em a}-Se$_{0.80}$S$_{0.20}$, but some features change as the concentration of 
sulphur increases. The peak around 234 cm$^{-1}$ (which is usually associated with Se chains \cite{Luo}) decreases and that at 
250 cm$^{-1}$ (associated to Se rings \cite{Luo}) increases indicating 
structural modifications on Se-Se units as sulphur is introduced on crystalline Se chains. In addition, 
the large band seen around 460 cm$^{-1}$ in {\em a}-Se$_{0.90}$S$_{0.10}$ is located in 
{\em a}-Se$_{0.80}$S$_{0.20}$ at 490 cm$^{-1}$.

These modes also appear in other Se-based alloys \cite{KleGeSe,GaSeRaman,Sugai}. The more interesting modes, however, are 
those associated with the broad peak around 334 cm$^{-1}$ in {\em a}-Se$_{0.90}$S$_{0.10}$ 
and around 352 cm$^{-1}$ in {\em a}-Se$_{0.80}$S$_{0.20}$, which cannot be associated with Se-Se vibrations, either 
in crystalline or amorphous Se, or with S-S vibrations. Ward \cite{Ward} obtained RS measurements for 
Se$_{0.05}$S$_{0.95}$ and Se$_{0.33}$S$_{0.67}$ crystalline alloys and found three Se-S modes in this region, 
at 344 cm$^{-1}$, 360 cm$^{-1}$ and 380 cm$^{-1}$. 
Thus, the conclusion is that Se-S pairs are found in our alloy, besides Se-Se and S-S ones and, since the relative 
intensity of the modes in this region is increasing, we could expect an increase in the contribution of Se-S 
vibrations, and probably an increase in the Se-S average coordination number. By fitting the Raman 
peaks using Lorentzian functions we have determined the Raman modes given in ta\-ble~\ref{tabraman}.

\begin{table}[h]
\caption{\label{tabraman} Raman modes obtained from fitting the peaks seen at fig. \ref{fig2} to Lorentzian 
functions and their occurence in {\em c}-Se, {\em a}-Se and Se-S alloys.}
\begin{tabular}{ccc}\hline
Mode & Frequency  & Mode  \\
& (cm$^{-1}$) & found in   \\\hline
$A_1$, $E$ & 239  & {\em c}-Se and {\em a}-Se\\
$A_2$, $E$ & 253 & {\em c}-Se and {\em a}-Se \\
$A_1$ & 338 & {\em c}-Se$_{0.33}$S$_{0.67}$  \\
$A_1$ & 352 & {\em c}-Se$_{0.33}$S$_{0.67}$ and {\em c}-Se$_{0.05}$S$_{0.95}$  \\
$A_1$  & 445 & {\em c}-Se$_{0.33}$S$_{0.67}$ and {\em c}-Se$_{0.05}$S$_{0.95}$ \\
 & & and also {\em c}-Se and {\em a}-Se  \\
$A_1$ & 465 & {\em c}-Se$_{0.33}$S$_{0.67}$ and {\em c}-Se$_{0.05}$S$_{0.95}$ \\
& & {\em c}-Se and {\em a}-Se \\
$A_1$ & 498 & {\em c}-Se$_{0.33}$S$_{0.67}$ and {\em c}-Se$_{0.05}$S$_{0.95}$ \\
& & {\em c}-Se and {\em a}-Se \\\hline
\end{tabular}
\end{table}

\subsection{RMC Simulations}

To obtain structural data as average coordination numbers and average interatomic distances, we made RMC 
simulations \cite{rmcreview,rmc++} using the total structure factor ${\mathcal S}(K)$ obtained 
from XRD measurements through eq. \ref{eqstructurefactor1} and also the EXAFS $\chi(k)$ signal at Se K edge. 
The first point is to determine the density of the alloy. 
To do that, we used the procedure suggested in Ref. \cite{Gereben}, and made several simulations for 
different values of the density $\rho$ keeping minimum distances and $\delta$ fixed (see eqs. \ref{eqchi} and 
\ref{eqchiexafs}). Thus, 
we chose the density that minimized the $\psi^2_{eq}$ parameter, which was $\rho = 0.030$ atm/\AA$^3$. 
It should be noted that the density found for {\em a}-Se$_{0.90}$S$_{0.10}$ had the same value at least considering 
the error bars. 
Then, we made several simulations changing the minimum distances between atoms, in order to find the best 
values for these parameters. All the simulations performed followed the procedure below

\begin{enumerate}
\item First, hard sphere simulation without experimental data were made to avoid possible memory effects of the 
initial configurations in the results. These simulations were run until reaching at least $3 \times 10^6$ 
accepted moves.

\item Next, simulations using the XRD experimental data were performed to obtain an initial convergence.

\item Finally, the EXAFS signal was added to the simulations. After reaching the convergence the 
statistically independent configurations were collected considering at least $1 \times 10^5$ accepted movements 
between one configuration and the next. At the end of the simulations, about 33\% of the generated 
movements were accepted.

\end{enumerate}

\begin{figure}[h]
\begin{center}
\includegraphics{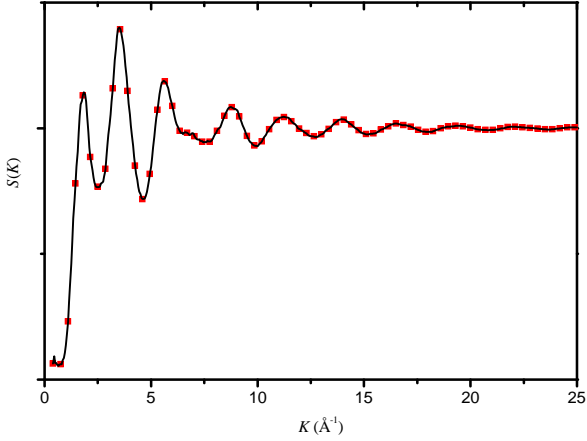}
\end{center}
\caption{\label{fig3} Experimental (full black line) and simulated 
(red squares) total structure factor for {\em a}-Se$_{0.80}$S$_{0.20}$.}
\end{figure}

The best values obtained for the minimum distances were $r_{min} = 2.20$ \AA\ for Se-Se and Se-S pairs and 
$r_{min} = 2.00$ \AA\ for S-S pairs. The experimental total structure factor ${\mathcal S}(K)$ and its RMC simulation 
obtained considering these values, 
16000 atoms and the density found are shown in fig. \ref{fig3}. As it can be seen, a very good agreement between 
them is reached.

\begin{figure}[h]
\begin{center}
\includegraphics{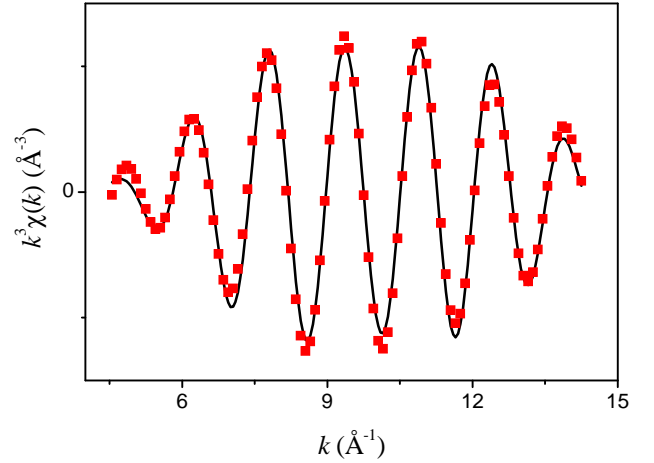}
\end{center}
\caption{\label{fig4} Experimental (full black line) and simulated 
(red squares) EXAFS $k^3\chi(k)$ signal at Se K edge for {\em a}-Se$_{0.80}$S$_{0.20}$.}
\end{figure}

Fig. \ref{fig4} shows the EXAFS $k^3\chi(k)$ signal and its RMC simulation. Again there is a good agreement between both 
functions. From the simulations, the partial distribution functions $g_{ij}(r)$ can be obtained and, from them, average 
coordination numbers and average interatomic distances can be determined. 

\begin{figure}[h]
\begin{center}
\includegraphics{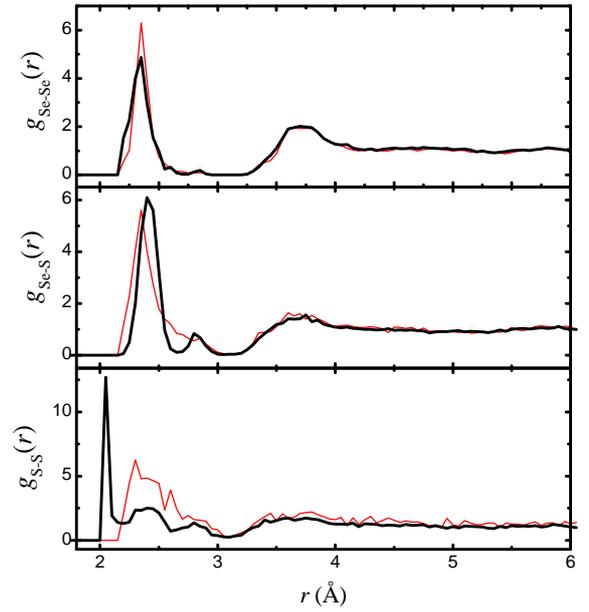}
\end{center}
\caption{\label{fig5} $g_{\text{Se-Se}}(r)$, $g_{\text{Se-S}}(r)$ and $g_{\text{S-S}}(r)$ 
functions obtained from the RMC simulations for {\em a}-Se$_{0.80}$S$_{0.20}$ (thick black lines)  and 
for {\em a}-Se$_{0.90}$S$_{0.10}$ (thin red lines) for comparison.}
\end{figure}

It is important to note that, 
concerning the XRD structure factor ${\mathcal S}(K)$, the main contribution to it comes from Se-Se pairs. For 
instance, at $K = 1$ \AA$^{-1}$, the coefficients $w_{ij}(K)$, given by eq. \ref{eqw}, are 
$w_{\text{Se-Se}} = 0.804$, $w_{\text{Se-S}} = 0.185$ and $w_{\text{S-S}} = 0.0107$. Thus, the contribution of 
${\mathcal S}_{\text{Se-Se}}(K)$ to ${\mathcal S}(K)$ is about 80.4\% (see eq. \ref{eqstructurefactor}), while the 
contribution of ${\mathcal S}_{\text{Se-S}}(K)$ is around 18.5\%, and the remaining 1.1\% is associated 
to ${\mathcal S}_{\text{S-S}}(K)$. This fact makes the error bars of the structural parameters related to 
Se-Se pairs much smaller than those associated with Se-S pairs and we have only estimates concerning S-S pairs.
Fig. \ref{fig5} shows the $g_{ij}(r)$ functions obtained from the RMC simulations for {\em a}-Se$_{0.80}$S$_{0.20}$ 
(thick lines) and for {\em a}-Se$_{0.90}$S$_{0.10}$ (thin lines), for comparison. 

Fig. \ref{fig5} shows interesting features. First, considering the $g_{\text{Se-Se}}(r)$ function, 
its maximum corresponding to the 
Se-Se first neighbors is located at $r^{\text{Se-Se}}_{max} = 2.35$ \AA, at the same place where it is found in 
{\em a}-Se$_{0.90}$S$_{0.10}$. However, the Se-Se average interatomic distance  is a little longer, being  
$\langle r^{\text{Se-Se}} \rangle = 2.37$ \AA. In addition, there is a decrease in the Se-Se average 
coordination number, from $\langle N^{\text{Se-Se}}\rangle =1.8$ in {\em a}-Se$_{0.90}$S$_{0.10}$ to 
$\langle N^{\text{Se-Se}}\rangle = 1.6$ in {\em a}-Se$_{0.80}$S$_{0.20}$, as we expected. To obtain these values 
we considered the range between 1.90 \AA\ and 3.00 \AA\ in the calculations.

Considering now the $g_{\text{Se-S}}(r)$ function, the first point to note is the appearence of a small second shell 
close to the first peak in $g_{\text{Se-S}}(r)$. This second shell could be already present in 
{\em a}-Se$_{0.90}$S$_{0.10}$ but it was not well defined there. The use of EXAFS data in the RMC simulations for 
{\em a}-Se$_{0.80}$S$_{0.20}$ resolved it better and we could calculate the average coordination number associated with 
both shells. Here we considered that the Se-S first shell is formed by these two subshells, so the Se-S average 
coordination number for first neighbors is the sum of the values found for each subshell. The first subshell 
corresponds to $\langle N^{\text{Se-S}}_1 \rangle = 0.51$, and the second subshell contributes with 
$\langle N^{\text{Se-S}}_2 \rangle = 0.08$ to the total value $\langle N^{\text{Se-S}} \rangle = 0.59$, which is 
almost twice larger than the value obtained for {\em a}-Se$_{0.90}$S$_{0.10}$. Even if we consider only the 
first Se-S subshell, the average coordination number is larger than that of {\em a}-Se$_{0.90}$S$_{0.10}$ by 67\%. The 
total Se average coordination number, given by $\langle N^{\text{Se}} \rangle = \langle N^{\text{Se-Se}} \rangle + 
\langle N^{\text{Se-S}} \rangle$, is then $\langle N^{\text{Se}} \rangle = 2.2$, larger than the value 
given by Heiba {\em et al}. \cite{Heiba} ($\langle N^{\text{Se}} \rangle = 2$) and by Shama \cite{Shama} 
($\langle N^{\text{Se}} \rangle = 2.1$).

The second point about $g_{\text{Se-S}}(r)$ is that the position of the first shell increased. The first 
Se-S subshell is found at $\langle r^{\text{Se-S}}_1 \rangle = 2.42$ \AA, and its 
maximum is at $r^{\text{Se-S}}_{1,max} = 2.40$~\AA, which is larger than the value found for 
{\em a}-Se$_{0.90}$S$_{0.10}$ in Ref. \cite{klese90s10} for the same shell ($\langle r^{\text{Se-S}} \rangle = 
2.34$ \AA). For the second subshell we found $\langle r^{\text{Se-S}}_2 \rangle = 2.83$ \AA\ and 
$r^{\text{Se-S}}_{2,max} = 2.80$~\AA. Thus, considering both subshells we have $\langle r^{\text{Se-S}} \rangle = 2.48$ 
\AA. The ranges used for the average calculations were [1.90--2.66] for the first subshell and [2.66--3.05] for the 
second. 

Finally, we can now discuss the $g_{\text{S-S}}(r)$ function. As we already pointed out, the contribution of S-S pairs 
to the XRD ${\mathcal S}(K)$ is very small, so we are mainly interested on its qualitative features. As fig. \ref{fig5} 
shows, one hypothesis is to think of the S-S first shell as formed by three subshells and to find the partial 
average coordination numbers and interatomic distances for each subshell. However, in this case we 
followed a different procedure, considering the first shell as ranging from 2.00 \AA\ to 3.10 \AA\ and, with this 
choice, we found $\langle N^{\text{S-S}}\rangle =1.2$ and $\langle r^{\text{S-S}} \rangle = 2.27$ \AA. There is 
a clear increase in the S-S average coordination number as the sulphur concentration goes from $x=0.10$ in 
{\em a}-Se$_{0.90}$S$_{0.10}$ to $x=0.20$ in {\em a}-Se$_{0.80}$S$_{0.20}$. It should be noted that these sulphur atoms 
belong to the amorphous Se$_{0.80}$S$_{0.20}$ phase, and the presence of a crystalline phase of sulphur can be ruled 
out since Raman results presented at fig. \ref{fig2} show the disappearance of the {\em c}-S Raman modes in both 
{\em a}-Se$_{0.90}$S$_{0.10}$ and {\em a}-Se$_{0.80}$S$_{0.20}$. All results above are summarized on 
table \ref{tab4}.

\begin{table}[h]
\caption{\label{tab4} Structural parameters obtained for {\em a}-Se$_{0.80}$S$_{0.20}$ from RMC simulations.}
\begin{tabular}{cccc}\hline
Bond type & Se-Se &  Se-S &  S-S \\\hline
$\langle N\rangle$  & $1.6 \pm 0.1$ & $0.59 \pm 0.1$ \footnotemark & $1.2 \pm 0.2$  \\
$\langle r\rangle$ (\AA)   & $2.37 \pm 0.05$  & $2.48 \pm 0.07$ \footnotemark & $2.27 \pm 0.15$  \\\hline
\end{tabular}

${}^2$ There are two subshells with $\langle N^{\text{Se-S}}_1 \rangle = 0.51$ and 
$\langle N^{\text{Se-S}}_2 \rangle = 0.08$, respectively.

${}^2$ There are two subshells at $\langle r^{\text{Se-S}}_1 \rangle = 2.42$ \AA\ and 
$\langle r^{\text{Se-S}}_2 \rangle = 2.83$~\AA, respectively.
\end{table}

\subsection{Optical Gap Energy Determination}

\begin{figure}[h]
\begin{center}
\includegraphics{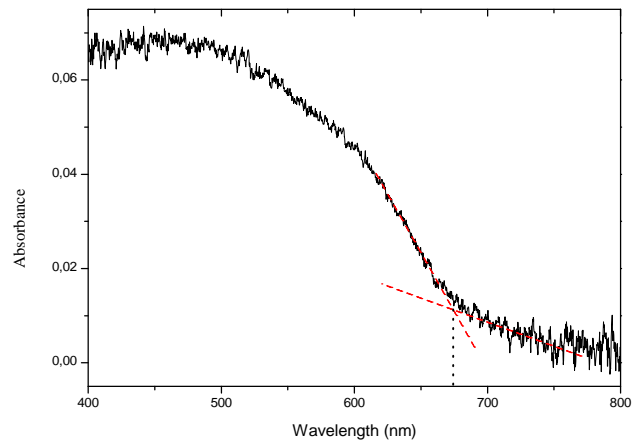}
\end{center}
\caption{\label{figabs} Absorption spectrum determined for {\em a}-Se$_{0.80}$S$_{0.20}$ and extrapolations to the 
baseline and to the absorption edge (dashed lines). The vertical dotted line indicates the crossing of both 
lines, which is the estimative of the band gap.}
\end{figure}

In addition to the structural and vibrational properties of {\em a}-Se$_{0.80}$S$_{0.20}$, 
we also calculated its optical 
gap energy. Fig. \ref{figabs} shows the absorbance obtained for this alloy. The absorption edge appears around 
610 -- 670 nm. As discussed on sec. \ref{secoptical}, we used two methods to estimate the 
optical gap. First, we made extrapolations of the baseline and the absorption edge \cite{Boldish}, which are shown in 
fig. \ref{figabs} (dashed lines). This procedure furnished a gap located at 674.4 nm, indicated in the figure by the 
vertical dotted line, corresponding to the value $E_g = 1.84$ eV.

\begin{figure}[h]
\begin{center}
\includegraphics{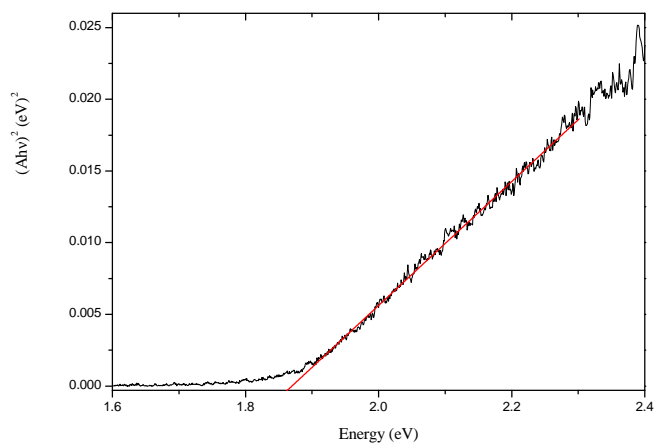}
\end{center}
\caption{\label{figgap} Plot of $(Ah\nu)^2 \times h\nu$ and linear fit for {\em a}-Se$_{0.80}$S$_{0.20}$ to obtain 
the optical gap energy.}
\end{figure}

We also used the McLean procedure \cite{McLean} to obtain the optical gap.  Fig \ref{figgap} shows the best fit 
achieved using eq. \ref{absorcao2}. We found a direct band gap ($n=2$) at 
$E_g = 1.87$ eV for {\em a}-Se$_{0.80}$S$_{0.20}$, indicating and increase in the gap energy with the 
increase in the sulphur concentration, since the value found for {\em a}-Se$_{0.90}$S$_{0.10}$ \cite{klese90s10} 
was $E_g = 1.81$ eV. For a comparison, Rafea and Farag~\cite{Rafea} obtained, for an 
amorphous Se$_{0.80}$S$_{0.20}$ thin film, $E_g = 1.96$~eV.

\section{Conclusion}

The amorphous Se$_{0.80}$S$_{0.20}$ alloy was prepared by MA and its local atomic structure 
investigated. The main conclusions of this study are:

\begin{enumerate}
\item The average number of Se-S pairs in the alloy increases as the sulphur concentration is raised from 
$x=0.10$ in {\em a}-Se$_{0.90}$S$_{0.10}$ to $x=0.20$ in {\em a}-Se$_{0.80}$S$_{0.20}$, 
showing that alloying occurs in an atomic level. Fukunaga {\em et al}. \cite{Fukunaga2} 
stated that there was only a mixing of Se chains and S rings in Se-S amorphous alloys, but Raman results and 
RMC simulations refuse this assumption. It is interesting to note that, although the Se-Se average coordination 
number decreased, the average total coordination number for Se atoms increases and is larger than 2.
In addition, the S-S average coordination number also increases as $x$ increases.

\item The density of the alloy seems to be independent of the sulphur concentration, at least for 
{\em a}-Se$_{0.90}$S$_{0.10}$ and {\em a}-Se$_{0.80}$S$_{0.20}$.

\item The optical gap of the alloy is also a function of the sulphur concentration, and it increases as the S 
content is increased, going from $E_g = 1.81$ eV for {\em a}-Se$_{0.90}$S$_{0.10}$ to 
$E_g = 1.87$ eV for {\em a}-Se$_{0.80}$S$_{0.20}$. 

\item Concerning the simulations, the use of EXAFS data on RMC simulations besides XRD ${\mathcal S}(K)$ produced 
better resolved shells, and furnished more reliable structural data.

\end{enumerate}

\begin{acknowledgement}

We thank the Brazilian agencies CNPq and CAPES for financial support. This study was partially 
supported by LNLS (proposal no 7093/08). We also thank Dr. Aldo J. G. Zarbin for help on Raman measurements.
\end{acknowledgement}


\end{document}